\documentclass[mathleft
]{an}
\usepackage{graphicx}
\usepackage{times}

\def\ltsima{$\; \buildrel < \over \sim \;$}
\def\simlt{\lower.5ex\hbox{\ltsima}}
\def\gtsima{$\; \buildrel > \over \sim \;$}
\def\simgt{\lower.5ex\hbox{\gtsima}}
\def\cgs{{erg cm$^{-2}$ s$^{-1}$}}

\def\cm2{{cm$^{-2}$}}

\def\p1{{Paper I}}

\def\xmm{{\em XMM--Newton}}
\def\chandra{{\em Chandra}}

\def\xmm{{\em XMM--Newton}}

\def\nh{{N$_{\rm H}$}}

\def\f14{{10$^{-14}$}}
\def\f13{{10$^{-13}$}}
\def\f12{{10$^{-12}$}}
\def\f11{{10$^{-11}$}}

\def\4u{{4U~1344$-$60}}

\def\nus{{\em NuSTAR}}

\def\aap{A\&A}

\begin{document}

% The following seven commands are intended for editorial usage and should be ignored by
% the author(s).
\Pagespan{1}{6}% Document's page range. 
% If second parameter is left empty, the last page is computed automatically.
\Yearpublication{}%
\Yearsubmission{}%
\Month{}%   
\Volume{}%  
\Issue{}% 
% \DOI{This.is/not.aDOI}% 

\title{The XMM Deep Survey in the CDFS}

\author{A. Comastri\inst{1}\fnmsep\thanks{Corresponding author:
  \email{andrea.comastri@oabo.inaf.it}\newline},
K. Iwasawa\inst{2,3},
C. Vignali\inst{4,1},
P. Ranalli\inst{5},
G. Lanzuisi\inst{4,1},
R. Gilli\inst{1}
%Example 
%for footnote, note the usage of the \texttt{fnmsep}
%command as separator between institute number and footnote mark} 
}
\titlerunning{XMM CDFS}
\authorrunning{A. Comastri et al.}
\institute{
INAF-Osservatorio Astronomico di Bologna, Via Ranzani 1, 40127, Bologna, Italy
\and
Institut de Ci\`encies del Cosmos (ICCUB), Universitat de Barcelona (IEEC-UB), Mart\`i i Franqu\`es, 1, 08028, Barcelona, Spain
\and
ICREA, Pg. Llu\`is Companys 23, 08010, Barcelona, Spain
\and
Dipartimento di Fisica e Astronomia, Universit\`a di Bologna, Viale Berti Pichat 6/2, 40127, Bologna, Italy
\and
Lund Observatory, Department of Astronomy and Theoretical Physics, Lund University, Box 43, 22100, Lund, Sweden
}

\received{}
\accepted{}
\publonline{later}


\abstract{%
  The Chandra Deep Field South (CDF--S) was observed by  \xmm\ for a total of
  about 3 Ms in many periods over the past decade (2001--2002 and 2008--2009). 
  The main goal of the survey was to obtain good quality X--ray
  spectroscopy of the AGN responsible for the bulk of the X--ray
  background. We will present the scientific highlights of the \xmm\
  survey and briefly discuss the perspectives of future observations
  to pursue XMM deep survey science with current and forthcoming X--ray facilities.}

\maketitle

\section{Introduction}

The primary goal of the XMM--{\it Newton} ultra--deep survey in the CDF--S 
was a detailed study of the X--ray spectral properties and cosmological evolution 
of heavily obscured AGN. 
The long exposure in the CDF--S, coupled with  the \xmm\ detector's
spectral throughput, made possible to 
obtain good quality ($> 10 \sigma$) spectra for some 170 X--ray sources ($\langle z
\rangle$=1.221; $\langle L_X \rangle$=4.8 $\times$ 10$^{43}$ erg
s$^{-1}$) over the range of redshifts and luminosities
which are relevant in terms of their contribution to the X--ray background. 

Since the early pioneering models (Setti \&  Woltjer 1989; Comastri
et al. 1995), it was realized that the XRB spectrum 
can be reproduced by an evolving population of unobscured and obscured
active galaxies (AGN) integrated over a broad range of
luminosities and up to very high redshift.
Obscured AGN  are known to produce the bulk  ($\sim$ 80\%)  of the
Universe accretion power and to play an important role in the
processes responsible for the
correlations between nuclear and host galaxy properties (Kormendy \&
Ho 2103). 
Gas and dust obscuration may represent a transient, but extremely important, phase in the
cosmic history of accretion onto  Supermassive Black Holes
(i.e. Hopkins et al. 2008).

From an X--ray perspective, obscured AGN are roughly subdivided in two
major classes depending if the  optical thickness for Compton
scattering of the obscuring gas is larger (lower) than unity. The
Compton thick threshold  (\nh$=\sigma_T^{-1}\sim1.5\times10^{24}$ cm$^{-2}$) 
corresponds to  an exponential cut--off in the X--ray
spectrum below approximately 10 keV. This value is coincident  with
the upper energy bound of focusing instruments of  large  X--ray
observatories, most notably XMM--{\it Newton} and {\it Chandra}.  
As a consequence, deep and ultra--deep surveys were extremely efficient
in the discovery and characterization of the Compton--thin
population up to high redshifts (Vito et al. 2014).
Furthermore, the much more elusive Compton thick AGN 
are a key  ingredient in all the models of the XRB and are required to fit the peak of its
emission at 20--30 keV and the source counts in the X--ray band.
Even though the number of bona--fide  Compton thick AGN has
significantly increased in the last decade (see the
review by Lanzuisi in this volume), their space density and
cosmological evolution remain poorly understood.

It turns out that excellent fits to the XRB spectrum and source
counts are obtained even for widely different assumptions on the spectra, space density and
evolution of the most obscured AGN (Gilli et la. 2007; Treister et
al. 2009; Ballantyne et al. 2011; Akylas et al. 2012; Ueda et al 2014;
Esposito \& Walter 2016)

The advent of \nus\,  with a spectral coverage extending up to
$\sim$ 80 keV and a sensitivity which has allowed to resolve about
$1/3$ of the 8--24 keV XRB (Aird et al. 2015; Harrison et al. 2016)
has started to improve our knowledge of high energy emission processes
in the most obscured AGN.
Despite the significant progresses obtained by the spectral analysis
of \nus\ survey sources (Zappacosta et al 2017; Del Moro et al. 2017),
deep \xmm\ and \chandra\ data are still extremely valuable 
to break the degeneracy between the various parameters.

Here we overview the results obtained by the \xmm\  deep survey in the
CDF--S (i.e. Comastri et al. 2011) in particular those related to the search for and the characterization 
of obscured AGN at cosmological distances.

\section{X-ray catalog}

The large effective area of \xmm\ in the 2--10 and 5--10 keV bands, coupled with
a 3.45 Ms nominal exposure time (2.82 and 2.45 Ms after light curve cleaning for MOS and PN, respectively), allowed us to build
clean samples in both bands, and makes the XMM-CDFS the deepest \xmm\ 
survey currently published in the 5--10 keV band. A three--band colour
image is shown in Figure~1.
A number of 339 and 137 sources are detected over an area of $\sim$
0.25 deg$^2$ in the bands reported above
with flux limits of 6.6$\times$ 10$^{-16}$ and 9.5$\times$ 10$^{-16}$ erg
cm$^{-2}$ s$^{-1}$ , respectively. The flux limits at 50\% of the maximum sky coverage are 
1.8$\times$ 10$^{-15}$ and 4.0$\times$ 10$^{-15}$ erg cm$^{-2}$
s$^{-1}$, respectively (Ranalli et al. 2013).
The cumulative flux of hard X--ray sources accounts for about 60\% of
the 5--10 keV XRB.

%%%%%%%%%%%%%%%%%%%%%%%%%%%%%%%%%%%%%%%%%%%%%%%%%%%%%%%%%%%%%%%%%%%%
\begin{figure}
\begin{center}
\includegraphics[width=7cm,height=7cm]{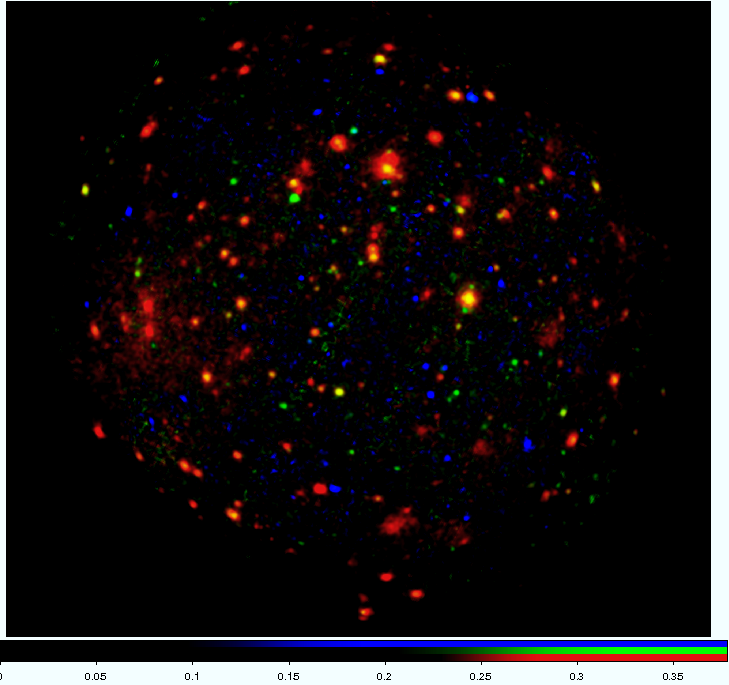}
\caption{Colour image of the XMM-CDFS. Red: 0.4--1  keV;  green:  1--2 keV;  blue:  2--8 keV.
The  colour  scaling  is  non-linear  (see Ranalli et al. 2013 for details).}
\label{fig:lumlum}
\end{center}
\end{figure}
%%%%%%%%%%%%%%%%%%%%%%%%%%%%%%%%%%%%%%%%%%%%%%%%%%%%%%%%%%%%%%%%%%%%%%%

\section{Spectral Analysis}

Above a limiting flux of the order of a few$\times10^{-15}$ \cgs\,
X--ray spectra obtained by \xmm\  long exposures have a number of counts
which allow us to constrain the column density and the Compton reflection
fraction, two key parameters of the XRB models.
The results of the spectral analysis of relatively bright, highly
obscured and Compton thick AGN  are presented  in Comastri et al. (2011),
Georgantopoulos et al. (2013) and Castello--Mor et al. (2013), while a
detailed analysis of the spectra and variability properties of the two
brightest source in the sample is reported in Iwasawa et al. (2015). 
Finally, a comprehensive analysis of the
X--ray spectra of a flux limited sample of  168 sources will be
presented in Iwasawa et al. (2017).

X--ray absorption is mainly measured by the low energy cut-off
of an X--ray spectrum, which moves to higher energies as the absorbing 
column density increases. When the the column density approaches the
Compton thick threshold  the cut--off is above 10 keV (i.e. outside
the \xmm\ band). 
 At the typical redshifts of the sources contributing to the
XRB ($z \sim$1), the cut--off,  along with the iron
$K\alpha$ complex and the reflection bump, are shifted toward lower energies, where the \xmm\ effective area is larger.

%%%%%%%%%%%%%%%%%%%%%%%%%%%%%%%%%%%%%%%%%%%%%%%%%%%%%%%%%%%%%%%%%%%%
\begin{figure}
\begin{center}
\includegraphics[width=7cm,height=6cm]{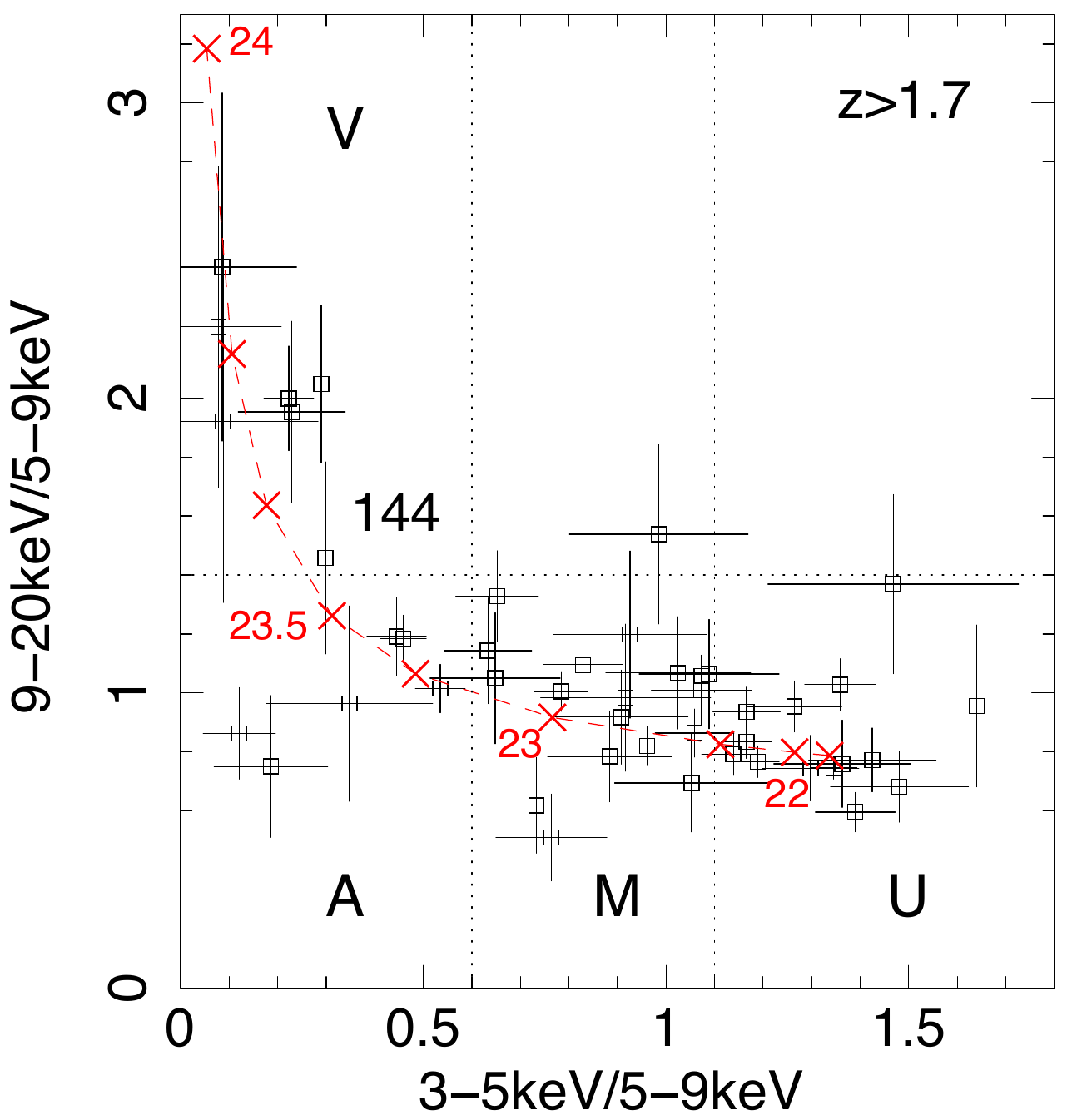}\vspace{0.1cm}
\includegraphics[width=7cm,height=6cm]{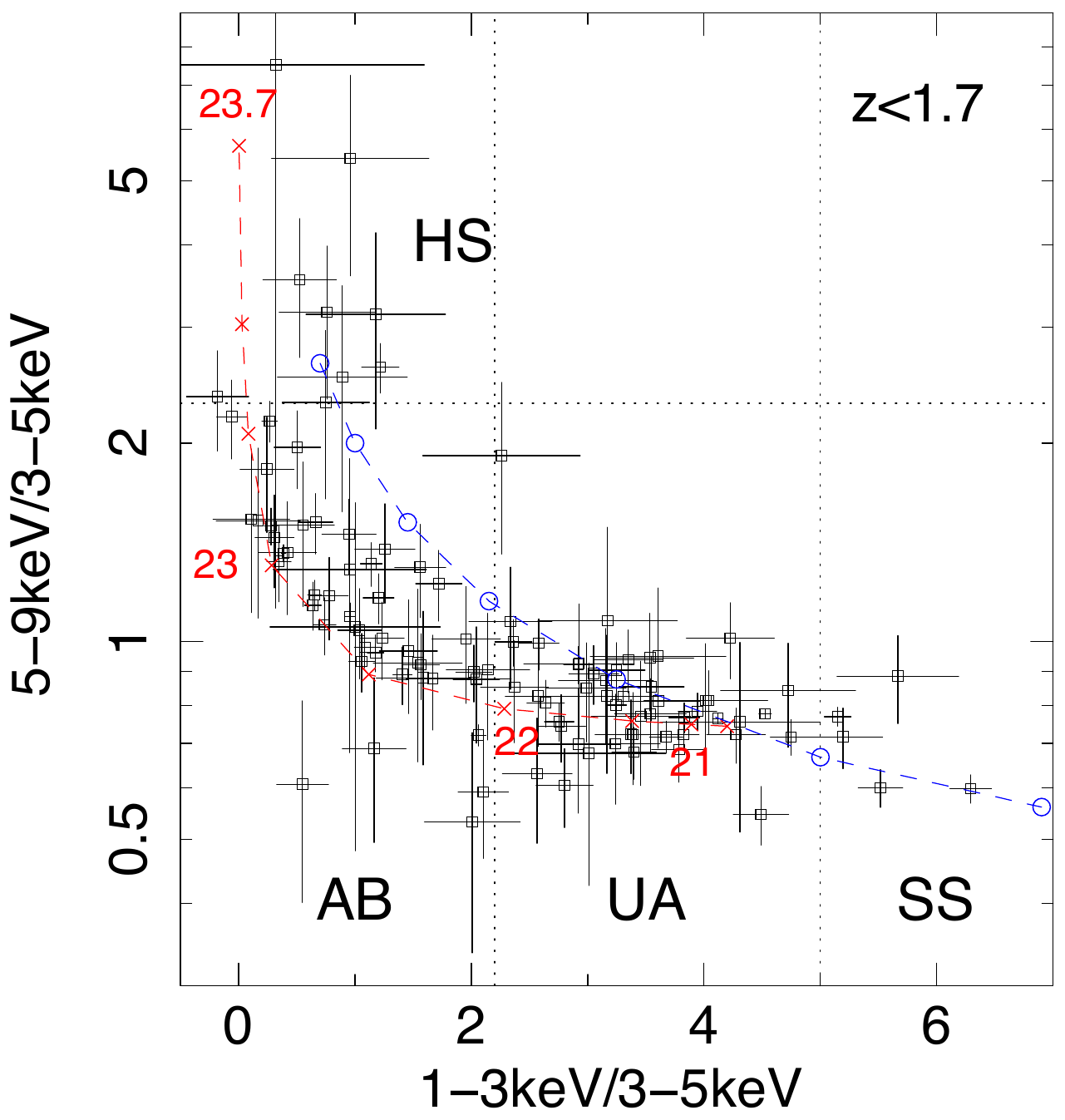}
\caption{{\it Top:} The color--color diagram for the high redshift ($z
  >$ 1.7) sources. Source 144 is the Compton thick quasar at z=3.7
  discussed in Comastri et al. (2011)  {\it Bottom:} The same for
  sources at $z< 1.7$. The dashed vertical and horizontal lines at
  fixed values of the hardness ratio subdivide the sample in four classes
for the stacking analysis. The red line, in both panels, represents
the locus of the hardness ratios 
corresponding to a power law continuum with $\Gamma=$1.8 modified by
absorption as labeled by the logarithm of the column density.}
\label{fig:ki1}
\end{center}
\end{figure}
%%%%%%%%%%%%%%%%%%%%%%%%%%%%%%%%%%%%%%%%%%%%%%%%%%%%%%%%%%%%%%%%%%%%%%%

Iwasawa et al. (2012) searched for heavily obscured AGN at $z > 1.7$
using three rest-frame  energy  bands: {\it s} (3--5 keV);
{\it m} (5--9 keV); and {\it h} (9--20  keV),  and two X--ray  colours:
{\it s/m} and {\it h/m}. For the adopted rest--frame energy range,
these X--ray colours are sensitive to column densities larger than
$N_H \simeq$ 10$^{22}$ cm$^{-2}$ and up to the Compton thick
threshold. 
In Fig.~2 (top panel), a locus of spectral evolution when a power--law continuum of
photon index $\Gamma=1.8$ is modified by various absorbing column of log
$N_H$ between 21 and 24 (cm$^{-2}$) is shown.
In the bottom panel a similar plot is reported for lower redshift
sources using different, more appropriate, bands: {\it s} (1--3 keV);
{\it m} (3--5 keV); and {\it h} (5--9 keV),  and the same  X--ray colors.

Sources lying in different parts of the two diagrams have X--ray
colors corresponding to different values of the nuclear
obscuration. The Very absorbed (V), absorbed (A), moderately absorbed
(M) and unabsorbed (U) groups of the high redshift panel were extensively
discussed in Iwasawa et al. (2012). A similar scheme (HS, AB, UA and SS) is used for the lower redshift group. 

%%%%%%%%%%%%%%%%%%%%%%%%%%%%%%%%%%%%%%%%%%%%%%%%%%%%%%%%%%%%%%%%%%%%
\begin{figure}
\begin{center}
\includegraphics[width=7cm,height=6.5cm]{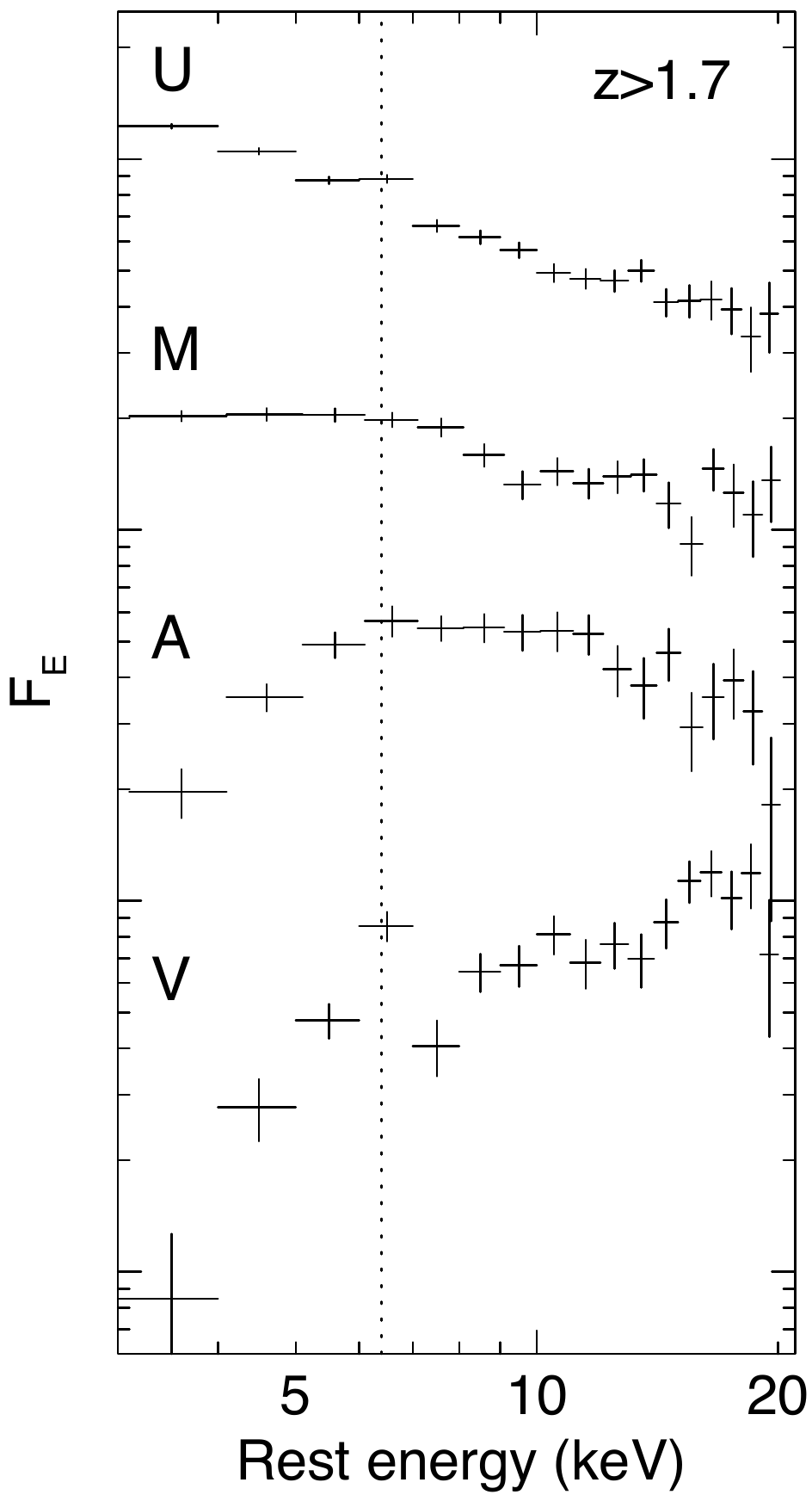}\vspace{0.1cm}
\includegraphics[width=7cm,height=6.5cm]{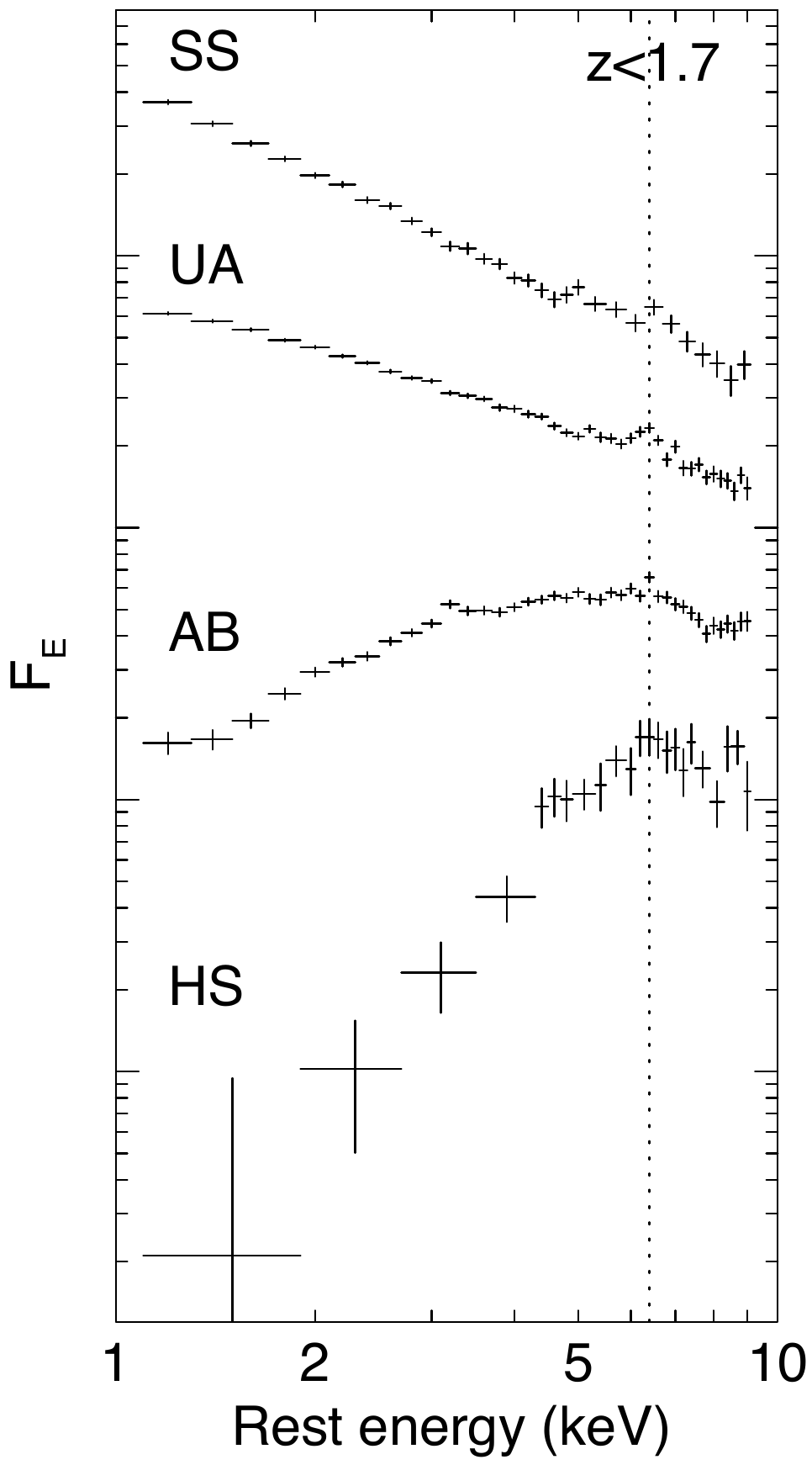}
\caption{The stacked spectra of the various sub classes of sources as
  labeled  above {\it top panel} and below {\it bottom panel} $z =
  1.7$. There are  46 sources in the high--$z$ sample and 110 in the
  low--$z$ sample. The bulk of the sources are in the M and U and 
the UA and AB classes. Compton thick AGN are mainly, but not uniquely
found in the V and HS classes.}
\label{fig:ki2}
\end{center}
\end{figure}
%%%%%%%%%%%%%%%%%%%%%%%%%%%%%%%%%%%%%%%%%%%%%%%%%%%%%%%%%%%%%%%%%%%%%%%

The average stacked spectra of the entire \xmm\ --CDFS spectral sample
grouped for different values of their rest--frame X--ray colors 
are reported in Figure~3 (both panels). The excellent spectral
quality,  especially at lower redshifts, is evident. 
The detailed spectral analysis of the most obscured sourecs in the HS
and V classes is currently undergoing. 
There is significant evidence of a redshift evolution of the fraction
of obscured AGN in high luminosity (\gtsima10$^{44}$ erg s$^{-1}$)  sources.
We anticipate that several previously unknown, or misclassified, Compton
thick AGN were identified and will be discussed in Iwasawa et
al. (2017).

\section{Iron features}

Evidence for ultra-fast (v $>$ 0.05 c) outflows in the innermost regions of AGN has been collected in the past
decade by sensitive X--ray observations for sizable samples of AGN,
mostly at low redshift. 
The \xmm\ spectrum of a luminous  ($L_{2-10 keV}$ $\sim$ 4 $\times$
10$^{44}$ erg s$^{-1}$), obscured ($N_H \sim 2 \times$ 10$^{23}$
cm$^{-2}$) quasar in the CDFS (PID 352) at $z = 1.6$ is shown in
Figure~4.
The source is characterized by an emission and absorption line complex 
in the 6--7 keV iron $K\alpha$ band. While the emission line is interpreted as being due to
neutral iron (consistent with the presence of cold absorption), the absorption feature is due to highly ionized iron transitions (FeXXV,
FeXXVI) with an outflowing velocity of 0.14$^{+0.02}_{-0.06}c$.
The mass outflow rate  is similar to the source accretion rate  ($\sim$ 2
M$_{\sun}$  yr$^{-1}$) , and the derived mechanical energy rate of almost
$10^{45}$ erg s$^{-1}$ corresponds to about 10\% of the source
bolometric luminosity. 
A highly ionized emission line is expected given the relatively large
solid angle $\Omega\sim 2\pi$ assumed to estimate the mass outflow
rate. The lack of detection may be due to a low  column
density for the ionized gas. As a consequence the emission feature 
intensity is not strong enough to be  detected in the \xmm\ spectrum.

PID352 represents one of the few cases where
evidences  of X--ray outflowing gas have been observed at high 
redshift thus far (Vignali et al. 2015).  
The relatively deep \xmm\ exposure -- $\sim$ 800 ks because the
source is at the edge of the field -- and the PID352 intrinsic
luminosity, made possible to collect enough source counts (4500 in
{\it EPIC} cameras) to detect the
outflowing X--ray feature.  The small area covered by the
\xmm\ deep field prevents a  systematic study of iron $K$ absorption features on
a statistically sound sample. 

%%%%%%%%%%%%%%%%%%%%%%%%%%%%%%%%%%%%%%%%%%%%%%%%%%%%%%%%%%%%%%%%%%%%
\begin{figure}[t]
\begin{center}
\hspace{0.7cm}
\includegraphics[width=6.5cm,height=8cm,angle=-90]{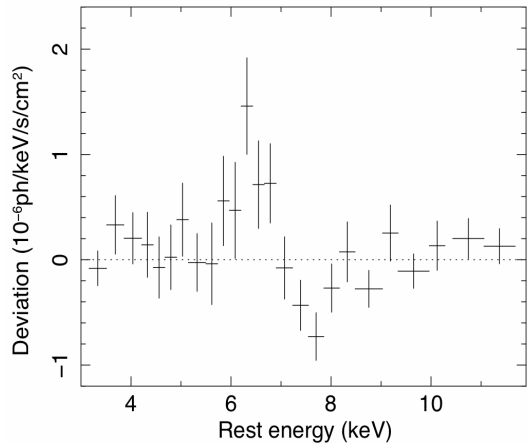}
\caption{The residuals with respect to an absorbed power law fit for
  the continuum of PID352 in the source rest frame. The iron $K\alpha$ emission
line, the neutral edge and the ionized absorption line at $\sim$ 7.5
keV are clearly visible.}
\label{fig:spec1}
\end{center}
\end{figure}
%%%%%%%%%%%%%%%%%%%%%%%%%%%%%%%%%%%%%%%%%%%%%%%%%%%%%%%%%%%%%%%%%%%%%%%

While the analysis of iron features in individual bright sources is
severely limited by the counting statistic, the stacking techniques do
not suffer from this limitation and return extremely useful
averaged information.
The spectral analysis of the 54 AGN  ($z_{median} \sim$ 1) with the best spectral
signal to noise ratio are presented in Falocco et al. (2013). Using a specifically
developed methodology to stack \xmm\ X--ray spectra they found a
robust ($\sim$ 7$\sigma$) and convincing evidence for the presence of iron line emission
up to $z\sim$ 3.5.
The average line equivalent width (EQW) is consistent with what is observed
in the local Universe for both obscured and unobscured
AGN. 
There is strong evidence of the so called Iwasawa--Taniguchi effect: the line EQW is much higher in lower luminosity Seyfert galaxies than in high
luminosity quasars  (Iwasawa \& Taniguchi 1993; Bianchi et al. 2007).  
The results of subdividing the CDF--S sample below and above a threshold
luminosity of 10$^{44}$ erg s$^{-1}$ are reported in Figure~5. 
The line intensity is clearly higher in lower luminosity Seyfert objects.

A combination of obscuration, enhancing the line
EQW and more common in lower luminosity AGN, and disk ionization,
quenching the line emission in luminous quasars, are  likely to
contribute to the observed effect.  The presence of an anti--correlation 
with the Eddington ratio (Bianchi et al. 2007) suggests that the accretion rate may be the
driver of the observed effect.

%%%%%%%%%%%%%%%%%%%%%%%%%%%%%%%%%%%%%%%%%%%%%%%%%%%%%%%%%%%%%%%%%%%%
\begin{figure}[t]
\begin{center}
\includegraphics[width=8cm,height=6cm]{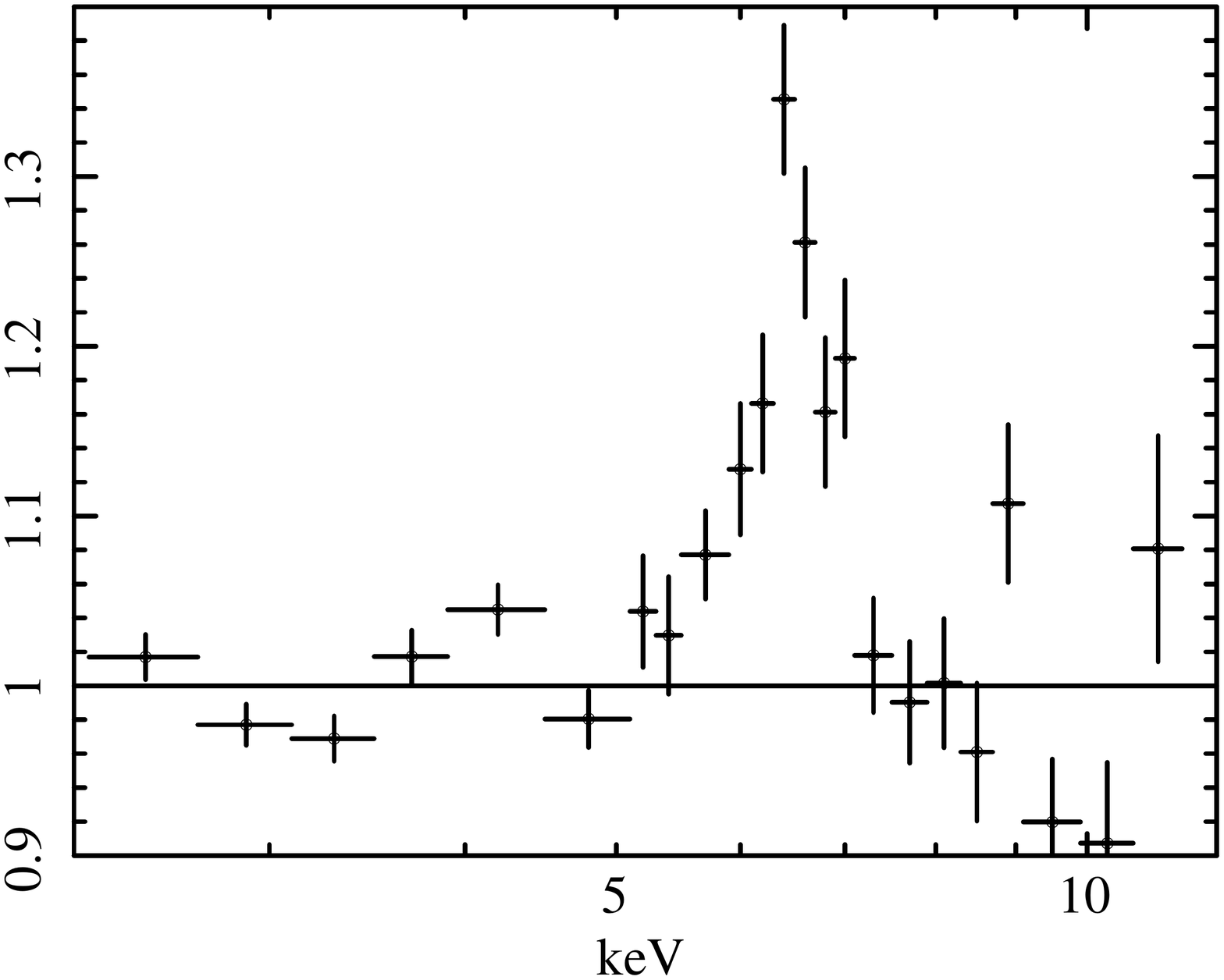}\vspace{0.1cm}
\includegraphics[width=8cm,height=6cm]{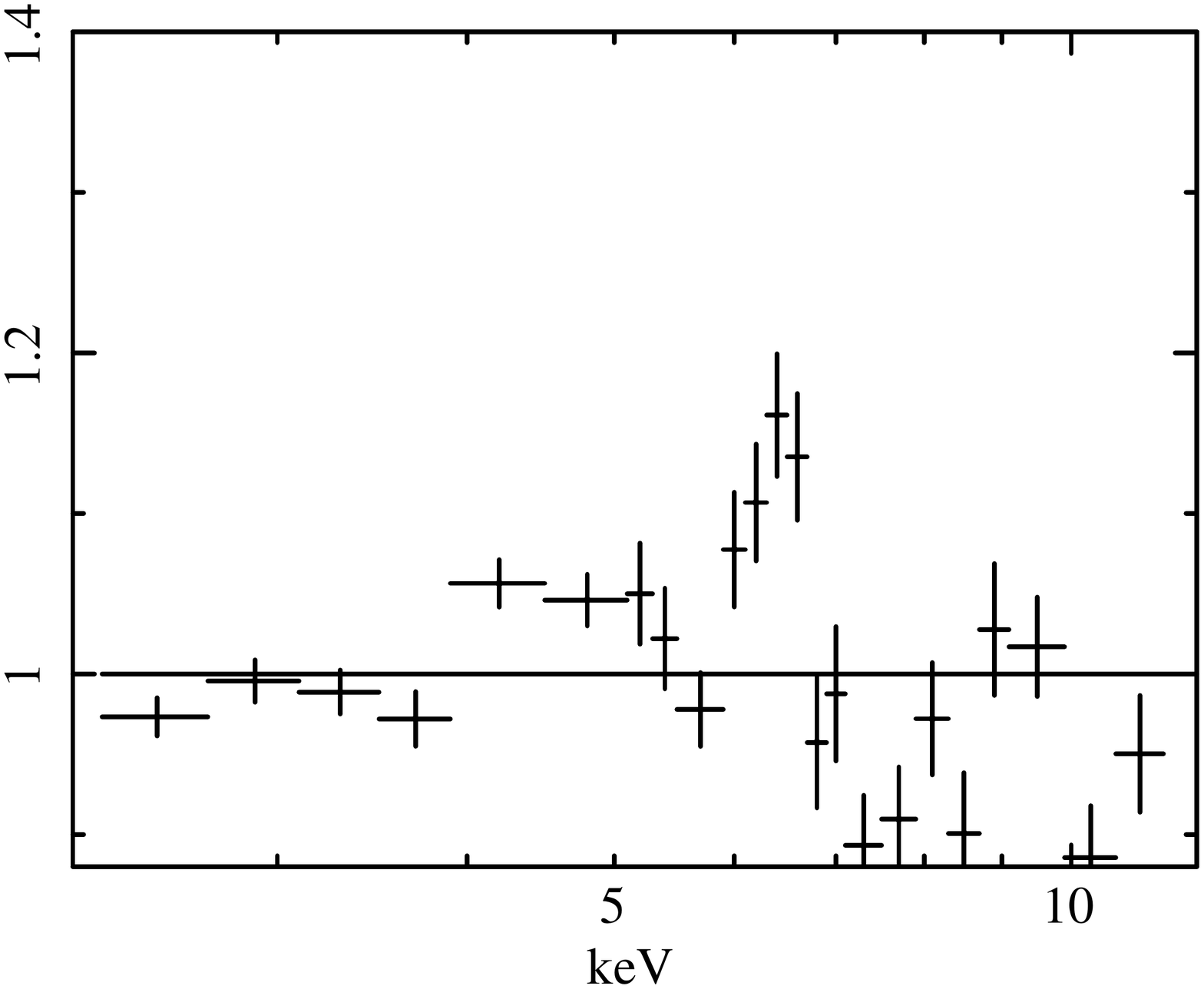}
\caption{{\it Top:} Residuals with respect to the best--fit power law
  of the stacked spectrum of the 33 objects in the low luminosity
  sub--sample at an average redshift $\langle z \rangle \simeq 0.9$ {\it Bottom:} As above but for the
 the high luminosity quasars sub--sample including 21 objects at an
 average redshift $\langle z \rangle \simeq 2.0$ }
\label{fig:spec2}
\end{center}
\end{figure}
%%%%%%%%%%%%%%%%%%%%%%%%%%%%%%%%%%%%%%%%%%%%%%%%%%%%%%%%%%%%%%%%%%%%%%%

\section{Conclusions}

The physics and evolution of the most obscured AGN is mainly based on
deep X--ray spectroscopy. Further progresses will be obtained
combining the data of  deep \xmm\ survey with  the ultra deep 7 Ms {\it Chandra}
observations in the CDF--S (Luo et al. 2016).  For the brightest sources the deep \nus\
survey in the same field has already extended the frequency coverage towards high
energies (Zappacosta et al. 2017). 

Current surveys are limited by both the relatively small field of view
covered in the CDF--S and, especially for \xmm\, the instrumental background. Additional deep
exposures will not allow to probe much deeper than the current limits
nor  to extend the source sample.

The rich plethora of multiwavelength data available in the deep fields
and, in particular MUSE and ALMA observations, will nicely complement
the deep X--ray spectroscopy and allow to further probe the physics of
obscured AGN.

A major advance in our understanding of the physics of the sources of
the XRB could be obtained with dedicated, deep  (of the order of 1
Ms each) exposures of carefully selected samples of sources such as
deeply buried AGN, strong iron line emitters and candidate
relativistic outflows.

\acknowledgements

The authors acknowledge financial support  from ASI-INAF grants
2015-046-R.0; I/037/12/10 and the PRIN-INAF 2014. We thank the member of the
XMM CDF--S Team and Massimo Cappi for interesting and extremely useful
discussions before the ``\xmm\ the next decade'' meeting.

\end{document}